\documentclass[12pt]{article}
\usepackage{graphicx}
\newcommand{\bb}{\begin{equation}}
\newcommand{\ee}{\end{equation}}
\newcommand{\ba}{\begin{eqnarray}}
\newcommand{\ea}{\end{eqnarray}}

\begin{document}

\title{{\bf Does Decoherence Make Observations Classical?}}

\author{
Don N. Page
\thanks{Internet address:
profdonpage@gmail.com}
\\
Department of Physics\\
4-183 CCIS\\
University of Alberta\\
Edmonton, Alberta T6G 2E1\\
Canada
}

\date{2021 August 30}

\maketitle
\large
\begin{abstract}
\baselineskip 20 pt

The fact that we rarely directly observe much quantum uncertainty is often attributed to decoherence.  However, decoherence does not reduce the quantum uncertainty in the full quantum state.  Whether or not it reduces the quantum uncertainties in observations depends on the yet-unknown rules for getting observations (and their measures or `probabilities') from the quantum state.  These points are illustrated by a simple toy model with a baseball at 100 miles per hour, which has the Planck momentum.

\end{abstract}

\normalsize

\baselineskip 22 pt

\newpage

\section{Introduction}

Decoherence is the development of quantum correlations between a quantum subsystem and its environment.  Since the entire universe (or multiverse if the full quantum state is that of a collection of sometime interacting sub-universes) is believed to be the smallest completely isolated system it contains, any subsystem smaller than the whole universe (or multiverse, though here I shall call the whole thing our ``universe'') has, at least at some time, interactions with other parts in its environment and hence can develop quantum correlations with its environment.

When any subsystem has quantum correlations with its environment, it cannot be in a pure state but must be in a mixed state.  That is, it cannot be correctly described by a wavefunction but instead only by a density matrix (for simplicity leaving aside such possibilities as a C*-algebra state).  In particular, there is no complete set of observables for which the subsystem has precise values as would be the case if the subsystem were in a pure state (given by a wavefunction) with no quantum correlations with its environment.  To put it another way, the density operator corresponding to the density matrix of the correlated subsystem is not a rank-one projection operator and hence is not an eigenstate of unit eigenvalue for any rank-one projection operator.

The density operator $\rho$ of a mixed state can be written as the sum of orthogonal rank-one projection operators ${\bf P}_i = |\psi_i\rangle\langle\psi_i|$ (the Gram-Schmidt basis \cite{Gram,Schmidt}) with nonnegative coefficients $p_i$ summing to unity that can be considered almost as if they were the probabilities for the full density operator to be one of the projection operators ${\bf P}_i$:  $\rho = \sum_i {p_i {\bf P}_i} = \sum_i {p_i |\psi_i\rangle\langle\psi_i|}$.

Then the von Neumann entropy of the subsystem with mixed state density operator $\rho$ is $S = - \mathrm{tr}\rho\ln{\rho} = -\sum_i p_i\ln{p_i}$, which is positive if the subsystem quantum state is not pure, necessarily the case if the subsystem has quantum correlations with its environment.

\newpage
One formulation of the second law of thermodynamics is that with a suitable division of the universe into subsystems, the quantum state of the universe is such that the sum of the von Neumann entropies of the subsystems increases with time.

\section{H.\ Dieter Zeh's Seminal Papers on Decoherence and Their Consequences}

In 1970, H.\ Dieter Zeh published a seminal paper on decoherence \cite{Zeh}.  This paper emphasized that the customary previous description of a measurement apparatus $M$ and a measured system $S$ as jointly being in a pure state of $M+S$ is invalid, as both of these parts also interact with their environment $E$, the rest of the universe.  So although $M+S+E$ make up a closed system (i.e., the whole universe) that may be assumed to evolve unitarily (leaving aside possibilities such as information loss in black hole formation and evaporation, for which there seems to be ever-growing doubts among theorists working on that problem), $M+S$ does not evolve unitarily, so that combined subsyatem is not a closed system and does not evolve unitarily to stay in a combined pure state even if it had started in a pure state.  In other words, $M+S$ becomes a mixed state.

In 1985, Erich Joos and Zeh wrote another highly influential paper \cite{JZ}.  Among many other things, this paper discusses an object of mass $m$ with density matrix $\rho(x,x',t)$ moving in one dimension in the environment of many small scatterers but otherwise freely, which they show obeys, to some approximation, the equation of motion (master equation)
\bb
\frac{\partial}{\partial t}\rho(x,x',t) = \frac{\hbar}{2 m}\left(\frac{\partial^2}{\partial x^2}-\frac{\partial^2}{\partial x'^2}\right)\rho - \Lambda(x-x')^2\rho,
\ee 
\bb
\Lambda = n \sigma v k^2/(8\pi^2),
\ee
where $n$ is the number density of scattering particles, $\sigma$ is the effective cross section of the scattering, $v$ is the mean relative velocity, and $k = p/\hbar$ is the rms wavenumber of the scattering particles with rms momentum $p$.

With $y = x - x'$ and $z = x + x'$ and the gaussian ansatz
\bb
\rho(y,z,t) = \exp{\{-[A(t)y^2 + iB(t)yz + C(t)z^2 + D(t)]\}},
\ee
the evolution equation becomes
\bb
\frac{\partial}{\partial t}\rho(y,z,t) = \frac{2i\hbar}{m}\frac{\partial^2}{\partial y \partial z}\rho - \Lambda y^2 \rho,
\ee
which leads to the coupled ordinary differential equations
\ba
dA/dt &=& 4\hbar AB/m + \Lambda,\\
dB/dt &=& 2\hbar (B^2 - 4AC)/m,\\
dC/dt &=& 4\hbar BC/m,\\
D &=& -\ln{(2\sqrt{C/\pi})}.
\ea

Letting $\tau \equiv \hbar t/m$ (which has dimension $L^2$), defining $\lambda \equiv 2\Lambda m/(3\hbar)$  (which has dimension $L^{-4}$), and using the notation $f' \equiv df/d\tau = (m/\hbar) df/dt$, the solution of
\ba
A' &=& 4AB + \Lambda m/\hbar = 4AB + (3/2)\lambda, \\
B' &=& 2(B^2 - 4AC), \ \mathrm{and} \\
C' &=& 4BC
\ea
is  \cite{JZ}
\ba
A &=& (2XX''-X'^2)/(8X), \\
B &=& -X'/(4X), \\
C &=& 1/(8X), \ \mathrm{and} \\
D &=& (1/2)\ln{(2 \pi X)},
\ea 
with
\ba
X &=& \lambda \tau^3 + a_2 \tau^2 + a_1 \tau + a_0, \\
a_0\! &=&\! 1/[8C(0)],\\
a_1\! &=&\! -B(0)/[2C(0)],\\
a_2\! &=&\! 2A(0) + [B(0)]^2/[2C(0)].
\ea

The position and momentum uncertainties for this solution are
\ba
(\Delta x)^2 &=& X = \lambda \tau^3 + a_2 \tau^2 + a_1 \tau + a_0, \\
(\Delta p/\hbar)^2 &=& (1/2)X'' = 3\lambda \tau + a_2.
\ea
Therefore, $a_0 = (\Delta x)_0^2$, the initial position variance, and $a_2 = (\Delta p/\hbar)_0^2$, the initial momentum variance divided by the square of Planck's reduced constant $\hbar$.  As a result,
\ba
(\Delta x)^2 &=& X = \lambda \tau^3 + (\Delta p/\hbar)_0^2 \tau^2 + a_1 \tau + (\Delta x)_0^2, \\
16a_1^2 &=& 4 (\Delta x)_0^2 (\Delta p/\hbar)_0^2 - A(0)/C(0),
\ea
and one must have $A \geq C$ for the density matrix to be positive, so {\it if} $4 (\Delta x)_0^2 (\Delta p/\hbar)_0^2$ takes its minimum value of 1 under the uncertainty principle, one must have $A(0) = C(0)$ (the condition for the initial density matrix to be pure) and hence $a_1 = 0$, and then $(\Delta x)_0^2$ is the only free initial condition, so
\bb
(\Delta x)^2 = \lambda \tau^3 + (\Delta p/\hbar)_0^2 \tau^2 + (\Delta x)_0^2 = \lambda \tau^3 + \tau^2/[4 (\Delta x)_0^2)] + (\Delta x)_0^2.
\ee

Assuming a minimum uncertainty initial state, the evolved state is
\ba
\rho &=& \sqrt{4C/\pi}\exp{\{-[A(x-x')^2 + iB(x^2-x'^2) + C(x+x')^2]\}}, \\
A &=& \frac{3\lambda^2\tau^4 +\lambda\tau^3/(\Delta x)_0^2 + 12\lambda(\Delta x)_0^2 \tau + 1}
          {8[\lambda\tau^3 + (1/4)\tau^2/(\Delta x)_0^2 + (\Delta x)_0^2]}, \\
B &=& - \frac{6\lambda\tau^2 + \tau/(\Delta x)_0^2}
          {8[\lambda\tau^3 + (1/4)\tau^2/(\Delta x)_0^2 + (\Delta x)_0^2]}, \\
C &=& \frac{1}{8[\lambda\tau^3 + (1/4)\tau^2/(\Delta x)_0^2 + (\Delta x)_0^2]}.\\
\ea         
The eigenvalues of this density matrix for all nonnegative integers $n$ are 
\bb
p_n = \frac{N^n}{(N+1)^{n+1}},
\ee
with the expectation value of $n$ being
\bb
N = \langle n \rangle = (1/2)(\sqrt{A/C}-1),
\ee
and the von Neumann entropy is
\ba
S &=& - \mathrm{tr}\rho\ln{\rho} = -\sum_n p_n\ln{p_n} = N\ln{(N+1)}-N\ln{N}\\
 &=& \ln{N} + (N+1)\ln{(1+1/N)} = \ln{N} + 1 + 1/(2N) + O(1/N^2)\\
 &=& (1/2)\ln{(A/C)} + 1 - \ln{2} - (1/6)(C/A) + O(C^2/A^2).
\ea

\section{Example of a Baseball}

Consider a baseball initially moving toward a batter at 100 miles per hour ($v_b = 44.704$ m/s, nearly the fastest that any pitcher can throw one), with a quantum amplitude to be batted back at the same speed, which would have a momentum of one Planck unit, $p_\mathrm{Pl} = \sqrt{\hbar c^3/G} = m_\mathrm{Pl}c = 6.524785(72)$ kg m/s, if the mass of the baseball were $m_b = m_\mathrm{Pl}c/v_b = 0.1459553(16)$ kg = 5.148421(57) ounces, which is just above the middle (59th percentile) of the allowed range for a baseball, between 5 and 5+1/4 ounces:           

{\bf A baseball at 100 miles per hour has the Planck momentum.}         
 
(This is the same momentum that our first granddaughter Lydia, born 2020 August 5 with mass 3.747 kg, would have had if she had fallen freely a distance of 0.154 m, 30\% of her birth length of 52 cm, but henceforth I shall instead consider the idealized motion of a baseball with initial momentum variance $(\Delta p)_0^2 = p_\mathrm{Pl}^2$.) 

Consider an idealized quantum state of a baseball moving in one spatial dimension ($x$) with zero mean momentum but initial rms momentum the Planck momentum $p_\mathrm{Pl}$ and rms velocity $v_b = 100$ mph = 44.704 m/s.  Suppose that initially it is in a pure gaussian state with minimum position uncertainty, $(\Delta x)_0 = l_\mathrm{Pl}/2$, half the Planck length.  Let the baseball have radius $r_b = 0.0369$ m (and hence cross-sectional area $\sigma_b = \pi r_b^2 = 4.28\times 10^{-3}$ m$^2$) and evolve while interacting with the sea level U.S. Standard Atmosphere with mean air molecular mass $m_a = 4.80965\times 10^{-26}$ kg, density $\rho_a = 1.2250$ kg/m$^3$, and temperature $T_a = 288.15$ K (and hence rms molecular velocity $v_a = \sqrt{3kT_a/m_a} = 498.144$ m/s).  Take an evolution time equal to the flight time over level ground $t_b$ of a baseball with no air friction leaving the bat with a speed $v_b$ at an angle of $45^\circ$ above the horizontal, so $t_b = \sqrt{2}v_b/g = 6.44675$ s.

Then
\ba
\tau &=& \hbar t_b/m_b \sim 5\times 10^{-33} \mathrm{m}^2 \sim 2\times 10^{37}l_\mathrm{Pl}^2,\\
\lambda &\equiv& 2\Lambda m_b/(3\hbar) = m_b \sigma_b m_a \rho_a v_a^3/(3 h^3) 
\sim 3\times 10^{79} \mathrm{m}^{-4} \sim 2\times 10^{-60} l_\mathrm{Pl}^{-4},\\
l_\mathrm{Pl} &\equiv& \sqrt{\hbar G/c^3} = 1.616255(18)\times 10^{-35}\,\mathrm{m}.
\ea

After $t_b = \sqrt{2}v_b/g = 6.44675$ s, the idealized density matrix of the baseball in Planck units has $\tau \sim 2\times 10^{37}$ and $\lambda \sim 2\times 10^{-60}$ and is
\ba
\rho &=& \sqrt{4C/\pi}\exp{\{-[A(x-x')^2 + iB(x^2-x'^2) + C(x+x')^2]\}}, \\
X &=& (\Delta x)^2 = \lambda\tau^3 + \tau^2 + 1/4 \approx \tau^2 \sim 3\times 10^{74} \sim (300\ \mathrm{m})^2, \\
A &=& \frac{2XX''-X'^2}{8X} = \frac{3\lambda^2\tau^4+4\lambda\tau^3+3\lambda\tau+1}{8(\lambda\tau^3+\tau^2+1/4)} \approx \frac{1}{2}\lambda\tau \sim 2\times 10^{-23}, \\        
B &=& -\frac{X'}{4X} = -\frac{6\lambda\tau^2 + 4\tau}{8(\lambda\tau^3+\tau^2+1/4)} \approx -\frac{1}{2\tau} \sim -3\times 10^{-38}, \\
C &=& \frac{1}{8X} = \frac{1}{8(\lambda\tau^3+\tau^2+1/4)} \approx \frac{1}{8\tau^2} \sim 4\times 10^{-76}, \\
(\Delta p)^2 &=& \frac{1}{2}X'' = 3\lambda\tau + 1 \approx 1 = p_\mathrm{Pl}^2 = (6.524785(72)\ \mathrm{kg\, m/s})^2.
\ea

Note that the decohering interactions change the momentum variance by only $3\lambda\tau$, about one part in $10^{22}$, and dropping that gives
\bb
(\Delta x)^2 \approx (\Delta x)_0^2 + (\Delta p)_0^2 (t_b/m_b)^2 \approx (\Delta p)_0^2 (t_b/m_b)^2.
\ee

Joos and Zeh \cite{JZ} give the eigenstates of the density matrix as
\bb
\phi_n(x) = \frac{(AC)^{1/4}}{2^{n-1}n!\sqrt{\pi}} H_n[2(AC)^{1/4}x] \exp{\{-[2\sqrt{AC}+iB]x^2\}}
\ee
(with Hermite polynomials $H_n$), nearly the same as harmonic oscillator energy eigenstates with mass $m_b$ and oscillator period $P = 2\pi/\omega \approx 2\pi m_b\sqrt{\tau/\lambda}/(\hbar\, l_\mathrm{Pl}) \sim 200\,000$ years for the baseball, except for the extra phase $-Bx^2$ that does not change the probability density $|\phi_n(x)|^2$ but does greatly increase the expectation value of the energy for $\lambda\tau(\Delta x)_0^2 \ll 1$, as is the case.         

The expectation value of the `excitation' number $n$ for the baseball is $N = \langle n\rangle = (1/2)(\sqrt{A/C}-1) \approx  (1/2)\sqrt{A/C} \approx \sqrt{\lambda\tau^3}/l_\mathrm{Pl} = \sqrt{2\sqrt{2}\pi r_b^2 v_b^5 m_a \rho_a v_a^3/(3 h^2 g^3)} \sim 10^{26}$, which gives the eigenvalues of the density matrix of the idealized single-particle baseball as
\bb
p_n = N^n/(N+1)^{n+1} = (N+1)^{-1}e^{-n\ln{(1+1/N)}} \approx N^{-1}e^{-n/N}
\ee
and von Neumann entropy (ignoring internal thermal motion)
\bb
S = (N+1)\ln{(N+1)} - N\ln{N} \approx \ln{N} + 1 \approx 61.
\ee

The mean position for the $n$th eigenstate of the baseball density matrix is zero, $\langle x_n \rangle = 0$, but the variance of the position is         
\ba
\langle \Delta x_n^2 \rangle \!\!\!\!&=&\!\!\!\! \frac{2n+1}{8\sqrt{AC}}
= \frac{(\Delta x)_0^2 \lambda \tau^3 + (1/4) \tau^2 + (\Delta x)_0^4}
{\sqrt{3(\Delta x)_0^4\lambda^2\tau^4 + (\Delta x)_0^2\lambda\tau^3 + 12(\Delta x)_0^6\tau+(\Delta x)_0^4}}
(2n+1)\\
\!\!\!\!&\approx&\!\!\!\! \sqrt{\frac{\tau}{\lambda}}\, \frac{2n+1}{4(\Delta x)_0}
\approx \sqrt{\frac{6\pi \hbar^2 v_b^2 t_b}{r_b^2 m_a \rho_a v_a^3}}\, (2n+1)
\sim (2\times 10^{-11}\ \mathrm{m})^2 (2n+1).
\ea
This is very small for $n=0$, the eigenstate with the largest eigenvalue, $p_0 = 1/(N+1) \sim 10^{-26}$, but the weighted variance over all eigenstates is
\ba
\langle \Delta x_n^2 \rangle \!\!\!\!&=&\!\!\!\! \frac{\langle 2n+1 \rangle}{8\sqrt{AC}} = \frac{2N+1}{8\sqrt{AC}} = \frac{1}{8C} = \lambda \tau^3 + (1/4) \tau^2/(\Delta x)_0^2 + (\Delta x)_0^2\\
\!\!\!\!&=&\!\!\!\! \lambda \tau^3 + (\Delta p)_0^2 (t/m)^2 + (\Delta x)_0^2
\approx (\Delta p)_0^2 (t/m)^2 = (v_b t_b)^2 \approx (288\ \mathrm{m})^2.
\ea
the position variance of the full state.  Thus the eigenstates of the decohered density matrix do not explain observations with relatively small quantum uncertainties in the position (e.g., much less than the 288-metre rms uncertainty in the decohered state).

For $x \sim (\Delta x)_0 \approx 288$ m, the rms position uncertainty, the eigenstate phase $-Bx^2$ changes by a phase of the order of $2\pi$ in a time $\sim h/(0.5 m_b v_b^2) \sim 5\times 10^{-36}$ seconds, so one idea is to average the baseball density matrix over a time long compared with this time (but short compared with the baseball motion time $t_b$, during which $A$ and $C$ would change by a significant fraction).  This effectively gets rid of the coefficient $B$, while leaving $A$ and $C$ unchanged, in the baseball density matrix, so that it becomes
\ba
\bar{\rho} &=& \sqrt{4C/\pi}\exp{\{-[A(x-x')^2 + C(x+x')^2]\}}, \\
A &\approx& \frac{1}{2}\lambda\tau \approx (7.62419\ \mathrm{m}^{-2})[t/(6.44675\ \mathrm{s})], \\
C &\approx& (\Delta x)_0^2/(2\tau^2) \approx (1.50501\times 10^{-6}\ \mathrm{m}^{-2})[t/(6.44675\ \mathrm{s})]^{-2},
\ea          
giving the same entropy $S \approx 61 + (2/3)\ln{[t/(6.44675\ \mathrm{s})]}$ and
$N = \langle n \rangle \approx \sqrt{\lambda \tau^3}/(2(\Delta x)_0) \approx 1.12538\times 10^{26}[t/(6.44675\ \mathrm{s})]^{3/2}$.

\section{Proposal for Small-Uncertainty Observations}

To avoid observations with large uncertainties within each observation, as have most of the eigenstates of the baseball density matrix $\rho$ (or even its average $\bar{\rho}$ over some relatively small time), one might propose that observations are represented by relatively localized operators $A_k$ that are inherent in the ultimate theory but independent of the quantum state density matrix $\rho$ of the universe. so that the measure (analogous to the probability) of the $k$th observation is tr$(A_k\rho)$, the expectation value of the operator $A_k$ in the quantum state of the universe.         

For example, in the one-dimensional toy example of an object with position $x$, these could be narrow gaussians centered on $x_k$,
\bb
A_k = N_k\exp{\{-[\alpha_k(x-x')^2 + \gamma_k(x+x'-2x_k)^2]\}}.
\ee
Then the measure of the $k$th observation would depend on the density matrix mainly in a localized region surrounding $x_k$, though it is another question of how well the content of the observation would reflect that narrow localization.

Of course, it is very na\"{\i}ve to suppose that the density matrix of a baseball directly leads to observations of it.  As emphasized in Quantum Darwinism \cite{Z2000,OPZ1,OPZ2,BZ1,BZ2,BZ3,Z1,PR2009,ZQZ2009,ZQZ2010,RZ2010,RZ2011,Z2,Z3,TYGDZ,Z4}, we generally do not interact directly with a system or object (such as a baseball), but mainly by observing a tiny fraction of the environment of photons that scatter off it.  These photons redundantly record information about pointer states \cite{Z1981} of the object, which presumably are much more localized than the generic eigenstates of the density matrix of the object.  Furthermore, certain properties of the photons (such as which photoreceptor cells in the retinas of one's eyes they excite) are recorded in the eyes and brain of the observer in neural systems that would also undergo decoherence.  Therefore, the fact that the simple model of the baseball decohered by molecules of the atmosphere does has a density matrix whose eigenstates are highly delocalized by no means suggests that visual observations of the baseball would show it similarly delocalized.  However, it does show that decoherence need not localize the eigenstates of an object's density matrix.

\section{Zurek on the Eigenstates of the Density Matrix}

In many of his papers on decoherence \cite{Z1982,Z1989,Z1993,Anderson,Z1991,PZ1993,ZP1995,Z1998a,Z1998b,Z2003a,Z2003b},\\
Wojciech Zurek and his colleagues have cautioned against taking the (Gram-Schmidt \cite{Gram,Schmidt}) eigenstates of the density matrix of the system as the observed pointer states \cite{Z1981}, though also noting varying degrees of similarity between them.
For example, in \cite{Z1982}, Zurek wrote, ``The resulting density matrix describing the correlated apparatus-system pair will be, to an excellent accuracy, diagonal in the pointer observable of the apparatus.''  Nevertheless, even if it is to high accuracy diagonal in the pointer observable, the basis in which the density matrix is precisely diagonal can be far from a spatially localized pointer basis, as shown by the example above of a baseball.

In \cite{Z1989}, William Unruh and Zurek wrote that the ``density matric decays rapidly toward a mixture of `approximate eigenstates' of the 'pointer observable'.''  In \cite{Z1993}, Zurek explained that ``diagonality alone is only a symptom---and not a cause---of the effective classicality of the preferred states,'' but instead focused on the ``ability to preserve correlations between the records maintained by the observer and the evolved observables of open systems as a defining feature of the preferred set of the to-be-classical states.''  My example of the baseball shows that diagonality in the pointer basis  \cite{Z1981} does not always occur, so it is a symptom that is not always present.

In response \cite{Anderson} to critiques of his 1991 {\it Physics Today} article \cite{Z1991}, Zurek wrote, ``It is clear from this recapitulation of the decoherence process and its consequences that an interpretation based solely on the instantaneous eigenstates of the density matrix of a single system would be, at best, 
na\"{\i}ve.''  I agree, since in my example the instantaneous eigenstates were nearly all highly nonlocal and hence not what is expected for the pointer basis.

Juan Pablo Paz and Zurek \cite{PZ1993} wrote a longer explanation of the relationship between the (Gram-Schmidt \cite{Gram,Schmidt}) eigenstates of the density matric and the pointer states:
``\ldots the reduced density matrix can always be instantaneously diagonalized. Its eigenstates, which are sometimes called Schmidt states \cite{Albrecht} are not necessarily identical (or even approximately the same) as the pointer states: The two sets of states can be expected to coincide only when the decoherence process has been effective which, in turn, implies restrictions on the time scales. Thus, the time at which the Schmidt states are calculated must be larger than the typical decoherence time scale of the problem.  In that case the Schmidt states becomes independent of the details of the initial condition and coincide with the pointer projectors.''  In my baseball example, even long after the decoherence time scale, when the decoherence has been highly effective, nearly all of the Schmidt states are highly nonlocal and hence still not at all close to the pointer states.

Two years later, Zurek and Paz \cite{ZP1995} wrote that ``na\"{\i}ve attempts at an interpretation based solely on the instantaneous eigenstates of the density matrix are, at best, a poor caricature of the implications of envoronment-induced decoherence.''

In 1998, Zurek \cite{Z1998a} noted, ``The loss of purity and the simultaneous loss of information is caused by the decoherence which is eliminating entanglement (as it should, to bring about resemblance of classicality) but which is exacerbated by the mismatch between the Gramm-Schmidt basis in the post-entanglement $|\psi_{\cal{SM}}\rangle$, and the pointer states of $\cal{S}$ and $\cal{M}$.''  In my example there is indeed a huge mismatch, though I am not certain that this exacerbates the loss of purity and of information.

That same year, Zurek also wrote \cite{Z1998b}, ``Einselection determines that pointer states will appear on the diagonal of the density matrix of the system.\ \ldots\ This eventual diagonalization of the density matrix in the einselected basis is a byproduct, an important symptom, but not the essence of decoherence.''  My example seems to be a case in which the density matrix never becomes diagonal in the einselected basis.

Zurek wrote in 2003 \cite{Z2003a}, ``Schmidt states \ldots have been even regarded as `instantaneous pointer states.'"  Later in 2003, he warned in his review paper \cite{Z2003b},
``The density matrix of a single object in contact with the environment will always be diagonal in an (instantaneous) Schmidt basis. This instantaneous diagonality should not be used as the sole
criterion for classicality (although see Zeh, 1973, 1990;  Albrecht, 1992, 1993). Rather, the ability of certain states to retain correlations in spite of coupling to the environment is decisive.''  My baseball example seems to be a case in which eventual diagonality of the density matrix in the pointer basis is not only not a sufficient condition or criterion for classicality but also is not a necessary condition for classicality.

D.\ A.\ R.\ Dalvit, J.\ Dziarmaga, and Zurek \cite{DDZ2005} discuss four criteria of classicality and show that for an underdamped harmonic oscillator, pointer states are well defined, and all four criteria agree, unanimously giving coherent states.  They also give the example of ``a free particle undergoing quantum Brownian motion, for which most criteria select almost identical Gaussian states (although, in this case, the predictability sieve does not select well defined pointer states).''  However, they do not compare the Gaussian states with the Gram-Schmidt eigenstates of the free particle density matrix.

\section{Conclusions}

Decoherence can greatly increase the von Neumann entropy of the density matrix of a subsystem interacting with its environment, but, at least in simple examples such as the one above, it does not decrease the rms positional spread in the eigenstates of that density matrix.  Therefore, by itself it does not seem sufficient to explain the classicality of typical observations (the fact that they do not individually show the large quantum uncertainties that the quantum state usually has) if even well after the decoherence time scale one looks at the classicality properties of the eigenstates of the density matrix.          

H.\ D.\ Zeh's 1970 paper \cite{Zeh} noted, ``A theory of measurement must necessarily be empty if it does not have a substitute for psychophysical parallelism.''  Operators such as $A_k$, whose expectation values give the measures for observations as sentient experiences
\cite{P1994a,P1994b,P1995a,P1995b,P1996,P2001,P2011,P2017,P2020},
would be one such substitute.  However, we do not yet know what these operators are (though of course we would expect them to act directly on brain states rather than on the states of some observed object such as a baseball), so even if we did learn the so-called `theory of everything' for the complete dynamics of the universe, and even if we learned the quantum state of the universe, we would still not have a complete theory for the universe.

\section*{Acknowledgments}

This research was supported in part by the Natural Science and Engineering Council of Canada.  I am also grateful for discussions, both on Zoom and by email, with Wojciech Zurek, and for an email exchange with Maximilian Schlosshauer.

%\vspace{5cm}

\end{document}